\documentclass{SCGE}
\usepackage[dvipdfmx,bookmarks=true,colorlinks,%
citecolor=blue,linkcolor=blue,anchorcolor=blue,filecolor=blue,urlcolor=blue]{
hyperref}

\begin{document}
\bibliographystyle{sci-chin}

%%%%%新版式要加上这组
\begin{picture}(0,0){\rm
\put(0,-20){\makebox[160truemm][l]{\bf {\sanhao\raisebox{2pt}{.}}
Article  {\sanhao\raisebox{1.5pt}{.}}}}}
\put(0,-34){\jiuwuhao {\textcolor[rgb]{0.5,0.5,0.5}{\sf %Special Topic: Fluid Mechanics
}}}%%(11月注释：调\textcolor[rgb]{x,x,x}中的数字x越大越灰)
\end{picture}

\def\bm{\boldsymbol}

\def\dl{\displaystyle}
\def\du{\end{document}}
\def\d{{\rm d}}
\def\e{{\rm e}}
\def\i{{\rm i}}
\def\pi{{\uppi}}

% The author doesn't need fill in it.
\Year{2016} %
\Month{??} %
\Vol{??} %  卷号
\No{?} %  期号
\BeginPage{1} % 起页码
\AuthorMark{{\rm Zhang Z H}}  %(11月注释：页眉上的作者)
\AuthorMarkCite{{\rm Zhang Z H}. }
%(11月注释：citation中的作者)
\DOI{??} % The author doesn't need fill in it.
%\ArtNo{??}

% \title[short text for running head]{full title}{comments for title}
\title[Investigation of the two-quasiparticle bands in the doubly-odd
nucleus $^{166}$Ta using a particle-number conserving cranked shell model]
{Investigation of the two-quasiparticle bands in the doubly-odd
nucleus $^{166}$Ta using a particle-number conserving cranked shell model}

\author[1*]{ZHANG ZhenHua}{}
\footnote{*Corresponding author (email: zhzhang@ncepu.edu.cn)}%手动E-mail地址

\address[{\rm1}]{Mathematics and Physics Department,
              North China Electric Power University, Beijing 102206, China}

\maketitle \vspace{-3.5mm}{\footnotesize\begin{center}
%Received XX, 2016; accepted XX, 2016; published online XX, 2016
\end{center}}\vspace*{-5mm}

%     Abstract is required.
\begin{center}
\rule{16.5cm}{0.4pt}
\parbox{16.5cm}
{\begin{abstract}
The high-spin rotational properties of two-quasiparticle bands in the
doubly-odd ${}^{166}$Ta are analyzed using the cranked shell model
with pairing correlations treated by a particle-number conserving method,
in which the blocking effects are taken into account exactly.
The experimental moments of inertia and alignments
and their variations with the rotational frequency $\hbar\omega$
are reproduced very well by the particle-number conserving calculations,
which provides a reliable support to the configuration assignments
in previous works for these bands.
The backbendings in these two-quasiparticle bands are analyzed by the calculated
occupation probabilities and the contributions of
each orbital to the total angular momentum alignments.
The moments of inertia and alignments for the Gallagher-Moszkowski partners
of these observed two-quasiparticle rotational bands are also predicted.
\end{abstract}}
\end{center}\vspace*{-0.6cm}

\begin{center}
\parbox{16.5cm}
{\bf\jiuhao particle-number conserving method,
            pairing correlations,
            moment of inertia}%关键词
\end{center}

\begin{center}
{}
\CITA    %%(11月注释：Citation内容自动生成)
%\Cit{~~~???, et al. ???. Sci China-Phys Mech Astron, 2014, 57: 1--6, doi:}%%(11月注释：Citation内容需手动填写)
\end{center}

\textwidth=178truemm \textheight=236truemm%%%%%%新版式要加上

%%%%%%%%%%%%%%%%%%%%%%%%%%%%%%%%%%%%%%%%%%%%%%%%%%%%%%%%%%%%
\wuhao\vspace*{1.5mm}

\begin{multicols}{2}

%%%%%%%%%%%%%%%%%%%%%%%%%%%%%%%%%%%%%%%%%%%%%%%%%%%%%%%%%%%%
%% Text of article.
%%%%%%%%%%%%%%%%%%%%%%%%%%%%%%%%%%%%%%%%%%%%%%%%%%%%%%%%%%%%
%    Section headings
\renewcommand{\baselinestretch}{1.08} \baselineskip 12.2pt\parindent=10.8pt

\renewcommand{\thefootnote}

\section{Introduction}{\label{sec:intro}}

Compared to even-even and odd-$A$ nuclei,
the structure of doubly-odd nuclei is among the most complex
topics in nuclear physics because of the complexity of level structure
associated with contributions from both valence protons and neutrons.
However, they often provide a wealth of nuclear structure phenomena
such as Gallagher-Moszkowski (GM) splitting~\cite{Gallagher1958_PR0111-1282},
signature inversion~\cite{Kreiner1979_PRL43-1150, Kreiner1980_JPG6-L13}
and chiral structure~\cite{Frauendorf1997_NPA617-131}, {\it etc.},
which have been investigated both experimentally and theoretically~\cite{Boisson1976_PR26-99,
Bengtsson1984_NPA415-189, Hara1991_NPA531-221,
Liu1995_PRC52-2514, Liu1996_PRC54-719, Bark1997_PLB406-193, Jain1998_RMP70-843,
Xu2000_NPA669-119, Starosta2001_PRL86-971,
Olbratowski2004_PRL93-052501, Meng2010_JPG37-064025}.

Recently, a considerable amount of data of the high-spin rotational bands
in the odd-odd rare-earth nuclei, {\it e.g.}, $^{166, 168, 170}$Ta
\cite{Hartley2010_PRC82-057302, Wang2010_PRC82-034315, Aguilar2010_PRC81-064317},
$^{170, 172, 174, 176}$Re~\cite{Hartley2013_PRC87-024315, Hartley2014_PRC90-017301,
Guo2012_PRC86-014323, Cardona1999_PRC59-1298}, {\it etc.},
have been observed experimentally.
Light odd-odd Ta isotopes, which are characterized by
small quadrupole deformations, provide a good opportunity to our understanding
of the dependence of band crossing frequencies and angular momentum alignments on
the occupation of specific single-particle orbitals.
These data also provide an excellent testing ground
for various nuclear models, {\it e.g.}, the cranked Nilsson-Strutinsky
method~\cite{Andersson1976_NPA268-205},
the Hartree-Fock-Bogoliubov cranking model with Nilsson~\cite{Bengtsson1979_NPA327-139}
and Woods-Saxon potentials~\cite{Nazarewicz1985_NPA435-397, Cwiok1987_CPC46-379},
the projected shell model~\cite{Hara1995_IJMPE4-637},
the projected total energy surface approach~\cite{Tu2014_SCPMA57-2054},
the tilted axis cranking model~\cite{Frauendorf2001_RMP73-463},
the cranked relativistic~\cite{Afanasjev1996_NPA608-107}
and non-relativistic mean-field models~\cite{Dobaczewski1997_CPC102-166} {\it etc}.

It is well known that pairing correlations are very important in the low angular
momentum region, where they are manifested by reducing the moments of inertia (MOIs)
of the rigid-body estimation~\cite{Bohr1958_PR0110-936}.
Further investigation indicates that, at the high-spin region~($\hbar\omega \sim 0.8$~MeV),
although the MOIs of rotational bands tends to be the same with or without pairing interaction,
the backbending frequencies still have large differences~\cite{Liu2009_PRC80-044329}.
On the other side, due to the blocking effects, the MOIs of 1-quasiparticle (qp) bands in odd-$A$ nuclei
are usually larger than those of the ground state bands in adjacent even-even nuclei.
The blocking effects on MOIs of multi-qp bands are even more important.
Therefore, in order to understand the high-spin rotational property of one nucleus,
the pairing correlations and the blocking effects should be treated correctly.

In this paper, the cranked shell model (CSM) with
pairing correlations treated by a particle-number conserving
(PNC) method~\cite{Zeng1983_NPA405-1, Zeng1994_PRC50-1388}
is used to investigate the rotational bands in the doubly-odd nucleus ${}^{166}$Ta.
In contrary to the conventional Bardeen-Cooper-Schrieffer or
Hartree-Fock-Bogolyubov approaches, in the PNC method, the Hamiltonian is
solved directly in a truncated Fock-space~\cite{Wu1989_PRC39-666}.
So the particle-number is conserved and the Pauli blocking effects are taken into
account exactly.
The PNC-CSM has already been employed successfully for describing odd-even differences
in MOIs~\cite{Zeng1994_PRC50-746},
the identical bands~\cite{Liu2002_PRC66-024320, He2004_CPL21-813, He2004_CPC28-1366, He2005_EPJA23-217},
the nonadditivity in MOIs~\cite{Liu2002_PRC66-067301, He2005_NPA760-263, Zhang2008_ChinPhysC32-681},
the nuclear pairing phase transition~\cite{Wu2011_PRC83-034323},
the rotational bands in the rare-earth~\cite{Liu2004_NPA735-77,
Zhang2009_NPA816-19, Zhang2009_PRC80-034313, Zhang2010_ChinPhysC34-39,
Zhang2010_ChinPhysC34-1836, Li2013_ChinPhysC37-014101, Zhang2016_NPA949-22},
the actinide and superheavy nuclei~\cite{He2009_NPA817-45,
Zhang2011_PRC83-011304R, Zhang2012_PRC85-014324, Zhang2013_PRC87-054308},
and the nuclear anti-magnetic rotation~\cite{Zhang2013_PRC87-054314}.
Note that the PNC scheme has been used both in relativistic
and nonrelativistic mean field models~\cite{Meng2006_FPC1-38, Pillet2002_NPA697-141} and
the total-Routhian-surface method with the
Woods-Saxon potential~\cite{Fu2013_PRC87-044319, Fu2013_SCPMA56-1423}.
Very recently, the particle-number conserving method based on the
cranking Skyrme-Hartree-Fock model has been developed~\cite{Liang2015_PRC92-064325}.

This paper is organized as follows.
A brief introduction to the PNC treatment of pairing correlations within
the CSM is presented in Sec.~\ref{sec:pnc}.
This method is used to investigate the 2-qp rotational
bands of ${}^{166}$Ta in Sec.~\ref{sec:resu}.
A brief summary is given in Sec.~\ref{sec:summ}.

\section{PNC-CSM formalism}{\label{sec:pnc}}

The CSM Hamiltonian of an axially symmetric nucleus in the rotating
frame can be written as
\begin{eqnarray}
 H_\mathrm{CSM}
 & = &
 H_0 + H_\mathrm{P}
 = H_{\rm Nil}-\omega J_x + H_\mathrm{P}
 \ ,
 \label{eq:H_CSM}
\end{eqnarray}
where $H_{\rm Nil}$ is the Nilsson Hamiltonian~\cite{Nilsson1969_NPA131-1},
$-\omega J_x$ is the Coriolis interaction with cranking frequency $\omega$ about the $x$
axis (perpendicular to the nuclear symmetrical $z$ axis), $H_0=H_{\rm
Nil}-\omega J_x$ is the one-body part of $H_{\rm CSM}$, and
$H_{\rm P}$ is the pairing interaction
\begin{eqnarray}
 H_{\rm P}
 & = &
  -G \sum_{\xi\eta} a^\dag_{\xi} a^\dag_{\bar{\xi}}
                        a_{\bar{\eta}} a_{\eta}
  \ ,
\end{eqnarray}
where $\bar{\xi}$ ($\bar{\eta}$) labels the time-reversed state of a
Nilsson state $\xi$ ($\eta$), and $G$ is the
effective strength of monopole pairing interaction.

Instead of the usual single-particle level truncation in conventional
shell-model calculations, a cranked many-particle configuration
(CMPC) truncation (Fock space truncation) is adopted which is crucial
to make the PNC calculations for low-lying excited states both
workable and sufficiently accurate~\cite{Wu1989_PRC39-666, Molique1997_PRC56-1795}.
Usually a dimension of 1000 should be enough for the calculations of the rare-earth nuclei.
An eigenstate of $H_\mathrm{CSM}$ can be written as
\begin{equation}
 |\Psi\rangle = \sum_{i} C_i \left| i \right\rangle
 \quad (C_i \; \textrm{real}) \ ,
\end{equation}
where $ |i\rangle $ is an CMPC (an eigenstate of the one-body operator $H_0$).
By diagonalizing the $H_\mathrm{CSM}$ in a sufficiently
large CMPC space, sufficiently accurate solutions for low-lying excited eigenstates of
$H_\mathrm{CSM}$ are obtained.

The angular momentum alignment for the state $| \Psi \rangle$ is
\begin{equation}
\langle \Psi | J_x | \Psi \rangle = \sum_i C_i^2 \langle i | J_x | i
\rangle + 2\sum_{i<j}C_i C_j \langle i | J_x | j \rangle \ ,
\end{equation}
and the kinematic MOI of state $| \Psi \rangle$ is
\begin{equation}
J^{(1)}=\frac{1}{\omega} \langle\Psi | J_x | \Psi \rangle \ .
\end{equation}
Because $J_x$ is a one-body operator, $\langle i | J_x | j \rangle$
($i\neq j$) may not vanish when two CMPCs $|i\rangle$ and
$|j\rangle$ differ by only one particle occupation. After a certain
permutation of creation operators, $|i\rangle$ and $|j\rangle$ can
be recast into
\begin{equation}
 |i\rangle=(-1)^{M_{i\mu}}|\mu\cdots \rangle \ , \qquad
|j\rangle=(-1)^{M_{j\nu}}|\nu\cdots \rangle \ ,
\end{equation}
where the ellipsis $\cdots$ stands for the same particle occupation,
and $(-1)^{M_{i\mu}}=\pm1$, $(-1)^{M_{j\nu}}=\pm1$ according to
whether the permutation is even or odd.
 Therefore, the angular momentum alignment of
$|\Psi\rangle$ can be expressed as
\begin{equation}
 \langle \Psi | J_x | \Psi \rangle = \sum_{\mu} j_x(\mu) + \sum_{\mu<\nu} j_x(\mu\nu)
 \ .
 \label{eq:jx}
\end{equation}
where the diagonal contribution $j_x(\mu)$ and the
off-diagonal (interference) contribution $j_x(\mu\nu)$ can be written as
\begin{eqnarray}
j_x(\mu)&=&\langle\mu|j_{x}|\mu\rangle n_{\mu} \ ,
\\
j_x(\mu\nu)&=&2\langle\mu|j_{x}|\nu\rangle\sum_{i<j}(-1)^{M_{i\mu}+M_{j\nu}}C_{i}C_{j}
  \quad  (\mu\neq\nu) \ ,
\end{eqnarray}
and
\begin{equation}
n_{\mu}=\sum_{i}|C_{i}|^{2}P_{i\mu} \ ,
\end{equation}
is the occupation probability of the cranked orbital $|\mu\rangle$,
$P_{i\mu}=1$ if $|\mu\rangle$ is occupied in $|i\rangle$, and
$P_{i\mu}=0$ otherwise.

The experimental kinematic MOI for each band is extracted by
\begin{equation}
\frac{J^{(1)}(I)}{\hbar^2}=\frac{2I+1}{E_{\gamma}(I+1\rightarrow
I-1)}
\end{equation}
separately for each signature sequence within a rotational band
($\alpha = I$ mod 2). The relation between the rotational frequency
$\omega$ and nuclear angular momentum $I$ is
\begin{equation}
\hbar\omega(I)=\frac{E_{\gamma}(I+1\rightarrow
I-1)}{I_{x}(I+1)-I_{x}(I-1)} \ ,
\end{equation}
where $I_{x}(I)=\sqrt{(I+1/2)^{2}-K^{2}}$, $K$ is the projection of
nuclear total angular momentum along the symmetry $z$ axis of an
axially symmetric nuclei.

\section{Results and discussion}{\label{sec:resu}}

In this work, the deformation parameters $\varepsilon_2= 0.192$ and $\varepsilon_4=-0.0045$
are taken from Ref.~\cite{Bengtsson1986_ADNDT35-15},
which are chosen as an average of the neighboring even-even Hf and W isotopes.
The Nilsson parameters ($\kappa$ and $\mu$) are taken as
the traditional values~\cite{Bengtsson1985_NPA436-14}
and a slight change for neutron $N=6$ major shell $\mu_6$ (modified from 0.34 to 0.28)
is made to account for the observed ground state in $^{166}$Ta.
In addition, the proton orbital $\pi 1/2^{-}[541]$ is slightly shifted upward by about 0.7~MeV,
which is adopted to avoid the defect caused by the velocity-dependent
$l^2$ term in the Nilsson potential for the MOIs and alignments
at the high-spin region~\cite{Andersson1976_NPA268-205}.
The effective pairing strengths can be determined by
the odd-even differences in nuclear binding energies,
and are connected with the dimension of the truncated CMPC space.
In this work, the CMPC space is constructed in the proton $N=4, 5$ shells
and the neutron $N=5, 6$ shells
with the truncation energies about 0.7$\hbar\omega_0$ both for protons and neutrons.
For $^{166}$Ta, $\hbar\omega_{\rm 0p}=7.160$~MeV
for protons and $\hbar\omega_{\rm 0n}=7.760$~MeV for neutrons~\cite{Nilsson1969_NPA131-1}.
The dimensions of the CMPC space are
about 1000 for both protons and neutrons in the calculation.
The corresponding effective pairing strengths are
$G_{\rm p}=0.34$~MeV for protons and $G_{\rm n}=0.46$~MeV for neutrons.
The stability of the PNC-CSM calculations against the change of the dimension of
the CMPC space has been investigated in
Refs.~\cite{Molique1997_PRC56-1795,Zeng1994_PRC50-1388}.
In the present calculations, almost all the
CMPCs with weight $>0.1\%$ are taken into account, so the solutions
to the low-lying excited states are accurate enough.
A larger CMPC space with renormalized pairing strengths gives essentially the same results.

\begin{figure}[H]
\centering
\includegraphics[width=1.0\columnwidth]{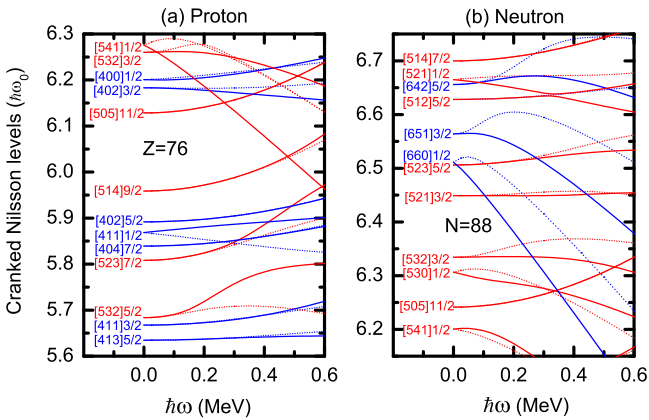}
\caption{\label{fig1} (Color online)
The cranked Nilsson levels near the Fermi surface of $^{166}$Ta
(a) for protons and (b) for neutrons.
The positive (negative) parity levels are denoted by blue (red) lines.
The signature $\alpha=+1/2$ ($\alpha=-1/2$) levels are denoted by solid (dotted) lines.
The deformation parameters $\varepsilon_2= 0.192$ and $\varepsilon_4=-0.0045$
are taken from Ref.~\cite{Bengtsson1986_ADNDT35-15},
which are taken as an average of the neighboring even-even Hf and W isotopes.
The Nilsson parameters ($\kappa$ and $\mu$) are taken as
the traditional values~\cite{Bengtsson1985_NPA436-14}
and a slight change for neutron $N=6$ major shell $\mu_6$ (modified from 0.34 to 0.28)
is made to account for the observed ground state in $^{166}$Ta.
In addition, the proton orbital $\pi 1/2^{-}[541]$ is slightly shifted upward by about 0.7~MeV.
}
\end{figure}

In Fig.~\ref{fig1}, the cranked Nilsson levels near the Fermi surface of $^{166}$Ta
(a) for protons and (b) for neutrons are shown.
The positive (negative) parity levels are denoted by blue (red) lines.
The signature $\alpha=+1/2$ ($\alpha=-1/2$) levels are denoted by solid (dotted) lines.
It can be seen that there exists a proton sub-shell at $Z=76$
and a neutron sub-shell $N=88$ near the Fermi surface.

\begin{figure}[H]
\centering
\includegraphics[width=0.8\columnwidth]{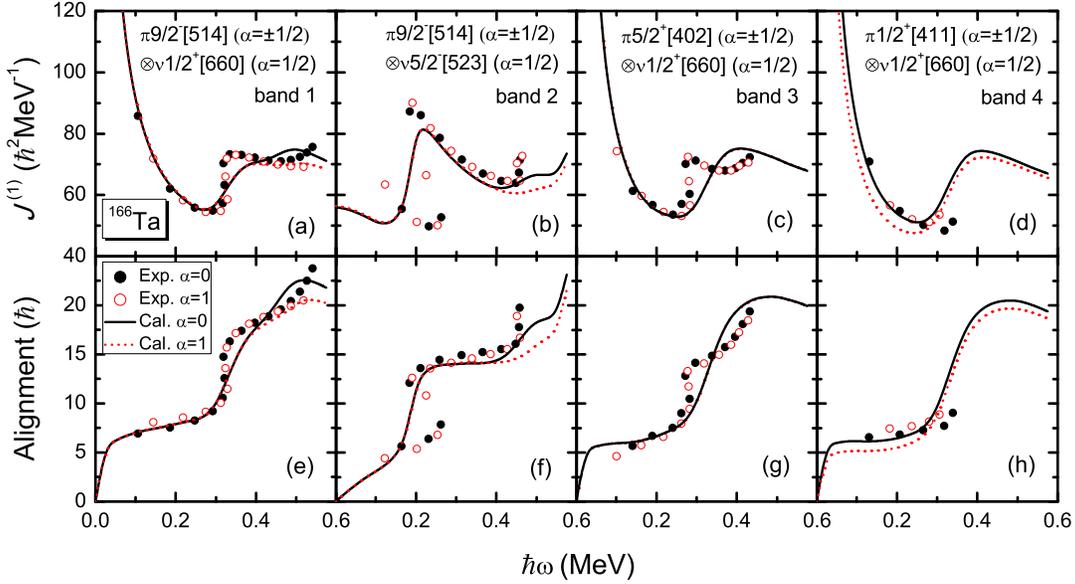}
\caption{\label{fig2} (Color online)
The experimental and calculated kinematic MOIs $J^{(1)}$ [(a)-(d)]
and alignments [(e)-(h)] of four low-lying bands in $^{166}$Ta.
The alignment $i$ is defined as $i= \langle
J_x \rangle -\omega J_0 -\omega ^ 3 J_1$ and the Harris parameters
$J_0 = 20\ \hbar^2$MeV$^{-1}$ and $J_1 = 40\ \hbar^4$MeV$^{-3}$ are
taken from Ref.~\protect\cite{Hartley2010_PRC82-057302}.
The experimental MOIs and alignments,
which are taken from Ref.~\cite{Hartley2010_PRC82-057302},
are denoted by black solid circles (signature $\alpha=0)$
and red open circles (signature $\alpha=1)$, respectively.
The calculated MOIs and alignments are
denoted by black solid lines (signature $\alpha=0$,
coupled with the neutron signature $\alpha=1/2$
and the proton signature $\alpha = - 1/2$) and red dotted lines
(signature $\alpha=1$, coupled with the neutron signature $\alpha=1/2$
and the proton signature $\alpha = 1/2$), respectively.
}
\end{figure}

Figure~\ref{fig2} shows the experimental and calculated kinematic MOIs $J^{(1)}$ (up row)
and alignments (bottom row) of four low-lying 2-qp bands in $^{166}$Ta.
The experimental data are taken from Ref.~\cite{Hartley2010_PRC82-057302}.
The alignment $i$ is defined as $i= \langle J_x \rangle -\omega J_0 -\omega ^ 3 J_1$
and the Harris parameters $J_0 = 20\ \hbar^2$MeV$^{-1}$ and $J_1 = 40\ \hbar^4$MeV$^{-3}$
are taken from Ref.~\cite{Hartley2010_PRC82-057302}.
The experimental MOIs and alignments are denoted by black solid circles
(signature $\alpha=0)$ and red open circles (signature $\alpha=1)$, respectively.
The calculated MOIs and alignments are
denoted by black solid lines (signature $\alpha=0$,
coupled with the neutron signature $\alpha=1/2$
and the proton signature $\alpha = - 1/2$) and red dotted lines
(signature $\alpha=1$, coupled with the neutron signature $\alpha=1/2$
and the proton signature $\alpha = 1/2$), respectively.
In Ref.~\cite{Hartley2010_PRC82-057302}, by analyzing the experimental alignments
and the electromagnetic transition probabilities of each rotational bands,
the configurations for these four 2-qp bands are assigned tentatively as
$\pi h_{11/2}(\alpha=\pm1/2) \otimes \nu i_{13/2}(\alpha=1/2)$ for band 1,
$\pi h_{11/2}(\alpha=\pm1/2) \otimes \nu f_{7/2}(\alpha=1/2) $ for band 2,
$\pi d_{5/2}(\alpha=\pm1/2)  \otimes \nu i_{13/2}(\alpha=1/2)$ for band 3 and
$\pi d_{3/2}(\alpha=\pm1/2)  \otimes \nu i_{13/2}(\alpha=1/2)$ for band 4,
which are labeled by the spherical quantum numbers.
For odd-odd nucleus, the total signature ($\alpha = 0, 1$) is coupled
by the odd proton ($\alpha = \pm 1/$2) and odd neutron ($\alpha = \pm 1/2$).
As for the yrast band in $^{166}$Ta [Fig.~\ref{fig2}(a)],
the configuration is assigned as $\pi h_{11/2} \otimes \nu i_{13/2}$.
The neutron $\nu i_{13/2}$ orbitals close to the Fermi surface, {\it i.e.},
$\nu 1/2^+[660]$ and $\nu 3/2^+[651]$, are all high-$j$ and low-$\Omega$ ones.
It can be seen from the cranked Nilsson levels in Fig.~\ref{fig1}(b) that,
both of these two orbitals have significant signature splitting,
while the experimental data in Fig.~\ref{fig2}(a) show very small signature
splitting at the low-spin region.
In addition, Fig.~\ref{fig1}(a) shows that the splitting of
the signature partners $\pi 9/2^-[514]$ $(h_{11/2})$ is very small.
So the signature partners in band 1 should be coupled from
$\alpha = +1/2$ (or $\alpha = -1/2$) of the odd neutron with
$\alpha =\pm 1/2$ of the odd proton to form the total signature $\alpha = 1, 0 $ (or 0, 1).
Moreover, for high-$j$ 1-qp configurations,
the favored signature is obtained by the simple rule,
$\alpha_{\rm f} = (-)^{j-1/2}1/2$~\cite{Stephens1975_RMP47-43}.
Therefore, it is reasonable to assign this configuration as
$\pi h_{11/2}(\alpha = \pm 1/2) \otimes \nu i_{13/2} (\alpha = 1/2)$.
The situations are similar for bands 2 and 4.
In the present calculation, I follow the configuration assignments in Ref.~\cite{Hartley2010_PRC82-057302}
and choose the corresponding configurations as
$\pi 9/2^-[514](\alpha=\pm1/2) \otimes \nu 1/2^+[660](\alpha=1/2)$ for band 1 [Fig.~\ref{fig2}(a)],
$\pi 9/2^-[514](\alpha=\pm1/2) \otimes \nu 5/2^-[523](\alpha=1/2)$ for band 2 [Fig.~\ref{fig2}(b)],
$\pi 5/2^+[402](\alpha=\pm1/2) \otimes \nu 1/2^+[660](\alpha=1/2)$ for band 3 [Fig.~\ref{fig2}(c)] and
$\pi 1/2^+[411](\alpha=\pm1/2) \otimes \nu 1/2^+[660](\alpha=1/2)$ for band 4 [Fig.~\ref{fig2}(d)],
which are denoted by the Nilsson quantum numbers.
It can be seen that after the configurations are determined,
the experimental MOIs and alignments of
all these four 2-qp bands can be reproduced quite well by the
PNC-CSM calculations, which in turn strongly support the
configuration assignments for these high-spin rotational bands
adopted in Ref.~\cite{Hartley2010_PRC82-057302}.
The signature splitting in Fig.~\ref{fig2}(a) after the first backbending is also
reproduced well by the present calculation.
It should be noted that the sharp backbendings appeared in the experimental MOIs
and alignments are not very well reproduced by the calculation.
This is because in the cranking model, before and after the backbending,
the two bands which have very different alignment from each other are mixed.
In order to obtain the backbending effect exactly, one has to go beyond the cranking model
and consider the two quasiparticle configurations in the vicinity of the
critical region~\cite{Hamamoto1976_NPA271-15, Cwiok1978_PLB76-263}.
Moreover, Fig.~\ref{fig2}(b) shows that the calculated second backbending frequency
($\hbar\omega \sim 0.55$~MeV) is a little larger than the data ($\hbar\omega \sim 0.46$~MeV).

\end{multicols}

\vspace*{-5mm}
\vspace*{-2mm}
\begin{multicols}{2}

\begin{figure}[H]
\centering
\includegraphics[width=1.0\columnwidth]{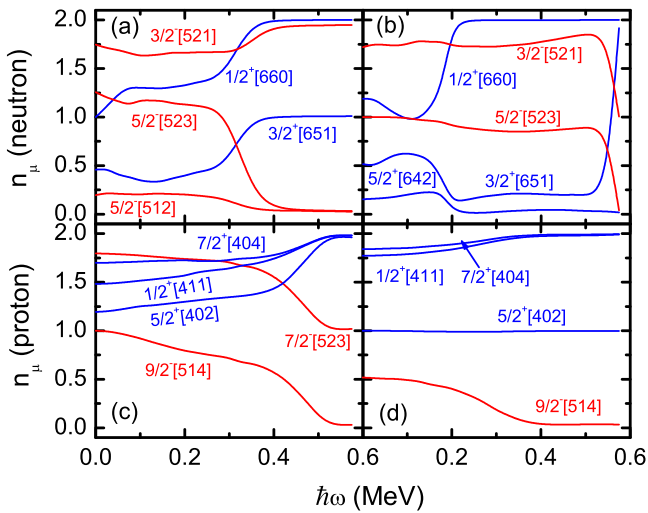}
\caption{\label{fig3} (Color online)
Occupation probability $n_\mu$ of each orbital $\mu$ near the Fermi surface
by blocking
(a) neutron $\nu 1/2^+[660](\alpha=1/2)$,
(b) neutron $\nu 5/2^-[523](\alpha=1/2)$,
(c) proton $\pi 9/2^-[514](\alpha=-1/2)$ and
(d) proton $\pi 5/2^+[402](\alpha=-1/2)$ in $^{166}$Ta.
The positive and negative parity levels are denoted by blue solid
and red dotted lines, respectively.
The Nilsson levels far above the Fermi surface
($n_{\mu}\sim0$) and far below ($n_{\mu}\sim2$) are not shown.
}
\end{figure}
\vspace*{-2mm}

One of the advantages of the PNC method is that the total particle
number $N = \sum_{\mu}n_\mu$ is exactly conserved from beginning to the end,
whereas the occupation probability $n_\mu$ for each orbital varies with
rotational frequency $\hbar\omega$.
By examining the $\omega$-dependence of the orbitals close to the Fermi surface,
one can learn more about how the Nilsson levels evolve with rotation and
get some insights on the backbending mechanism.
Figure~\ref{fig3} shows the occupation probability $n_\mu$ of each
orbital $\mu$ near the Fermi surface by blocking
(a) neutron $\nu 1/2^+[660](\alpha=1/2)$,
(b) neutron $\nu 5/2^-[523](\alpha=1/2)$,
(c) proton $\pi 9/2^-[514](\alpha=-1/2)$ and
(d) proton $\pi 5/2^+[402](\alpha=-1/2)$ in $^{166}$Ta.
The occupation probability of proton $\pi1/2^+[411]$ is not shown,
because it is nearly the same as that of the proton $\pi5/2^+[402]$ [Fig.~\ref{fig3}(d)].
The positive and negative parity levels are denoted by blue solid
and red dotted lines, respectively.
The Nilsson levels far above the Fermi surface
($n_{\mu}\sim0$) and far below ($n_{\mu}\sim2$) are not shown.
Because in the present PNC-CSM framework, the residual neutron-proton interaction is
not considered, the neutrons and the protons can be treated separately.
The combinations of the neutron [Fig.~\ref{fig3}(a) and (b)] and the proton
[Fig.~\ref{fig3}(c) and (d)] occupation probabilities can
account for the backbendings in the signature $\alpha = 0$ bands of band 1 to band 4.
It can be seen from Fig.~\ref{fig3}(a) that around the rotational frequency
$\hbar\omega \sim 0.30$~MeV, the occupation probabilities of
the first and the second lowest neutron $\nu i_{13/2}$ orbitals $\nu1/2^+[660]$ and $\nu3/2^+[651]$
increase sharply from about 1.3 to 2.0 and  0.5 to 1.0, respectively.
At the same time, the occupation probability of $\nu5/2^-[523]$ decreases from about 1.1 to nearly zero.
While in Fig.~\ref{fig3}(c) and (d), the occupation probability
of each proton orbital changes gradually around the first backbending region.
This indicates that the first backbendings ($\hbar\omega \sim 0.30$~MeV) observed in
band 1, band 3 and band 4 (data are not available in the backbending region)
may come from the contribution of the first and the second lowest $\nu i_{13/2}$ neutrons.
Figure~\ref{fig3}(b) shows that at $\hbar\omega \sim 0.20$~MeV,
the occupation probability of neutron $\nu1/2^+[660]$ increases from
about 1.0 to 2.0, while $\nu3/2^+[651]$ drops from 0.6 to 0.2.
At the same time, the occupation probability of $\nu5/2^+[642]$ drops from 0.2 to zero.
Therefore, the first backbending observed in band 2 may mainly come from the
contribution of $\nu1/2^+[660]$ and $\nu3/2^+[651]$, and the $\nu5/2^+[642]$
may also have a little contribution.
The alignment gain in signature $\alpha = 0$ of band 1 at higher
frequencies $\hbar\omega > 0.45$~MeV can be understood in Fig.~\ref{fig3}(c).
It can be seen that the proton occupation probabilities of
$\pi9/2^-[514]$ and $\pi7/2^-[523]$ drop quickly in this region,
so the alignment may come from these proton $\pi h_{11/2}$ orbitals.
The present calculation shows that in Fig.~\ref{fig3}(b),
the neutron orbital $\nu3/2^+[651]$ increase sharply from 0.2 to nearly 2.0.
Therefore, the second backbending ($\hbar\omega \sim 0.46$) in band 2
may come from the contribution of this orbital.
However, the PNC-CSM calculation overestimates this backbending frequency about 0.1~MeV.
If the single-particle levels were adjusted more carefully,
the calculated results may be improved further.
The mechanisms of the first and the second backbendings for these observed 2-qp bands in
the PNC-CSM calculation are consistent with those proposed in Ref.~\cite{Hartley2010_PRC82-057302}.
It also can be seen that when the high-$j$ orbital, {\it i.e.}, $\pi9/2^-[514]$ or $\nu1/2^+[660]$,
is blocked, a strong Coriolis mixing still exists between these high-$j$ orbitals.
In this case, the occupation probability of this high-$j$ orbital deviate from
$n_\mu \sim 1.0$ and there is no pure blocking.
While when the normal orbital is blocked, {\it i.e.}, $\pi 5/2^+[402]$ or $\nu 5/2^-[523]$,
the Coriolis mixing is negligible.

\vspace*{-2mm}
\begin{figure}[H]
\centering
\includegraphics[width=1.0\columnwidth]{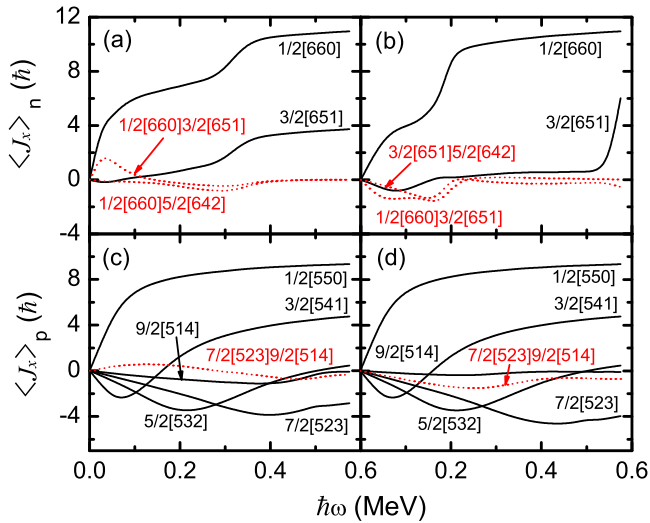}
\caption{\label{fig4} (Color online)
Contribution of each neutron orbital in the $N=6$
major shell (top row) and each proton orbital in the $N=5$
major shell (bottom row) to the angular momentum alignments $\langle J_x\rangle$ for
(a) neutron  $\nu 1/2^+[660](\alpha=1/2)$ ,
(b) neutron $\nu 5/2^-[523](\alpha=1/2)$,
(c) proton $\pi 9/2^-[514](\alpha=-1/2)$ and
(d) proton $\pi 5/2^+[402](\alpha=-1/2)$ in $^{166}$Ta.
The diagonal (off-diagonal) part $j_x(\mu)$
[$j_x(\mu\nu)$] in Eq.~(\protect\ref{eq:jx}) is denoted by black solid (red dotted) lines.}
\end{figure}
\vspace*{-2mm}

It is well known that the backbending in the rare-earth nuclei
mainly comes from the neutron $\nu i_{13/2}$ and proton $\pi h_{11/2}$ orbitals.
The calculated occupation probabilities in Fig.~\ref{fig3} also confirm this point.
In order to have a  more clear understanding of the backbending
mechanism, the contribution of each neutron orbital in the $N=6$
major shell (top row) and each proton orbital in the $N=5$
major shell (bottom row) to the angular momentum alignments
$\langle J_x\rangle$ for
(a) neutron  $\nu 1/2^+[660](\alpha=1/2)$ ,
(b) neutron $\nu 5/2^-[523](\alpha=1/2)$,
(c) proton $\pi 9/2^-[514](\alpha=-1/2)$ and
(d) proton $\pi 5/2^+[402](\alpha=-1/2)$ in $^{166}$Ta
are presented in Fig.~\ref{fig4}.
Note that in this figure, the smoothly increasing part of the
alignment represented by the Harris formula ($\omega J_0 +\omega^ 3 J_1$)
is not subtracted ({\it cf.} the caption of Fig.~\ref{fig2}).
The diagonal (off-diagonal) part $j_x(\mu)$ [$j_x(\mu\nu)$] in
Eq.~(\protect\ref{eq:jx}) is denoted by black solid (red dotted) lines.
In Fig.~\ref{fig4}(a), the PNC
calculation shows that around the backbending ($\hbar \omega \sim$~0.30 MeV) region,
the diagonal parts $j_x\left(\nu 1/2^+[660]\right)$ and
$j_x\left(\nu 3/2^+[651]\right)$ change a lot.
The alignment gain after the backbendings in band 1, band 2 and band 4
mainly comes these two terms.
At the same time, the off-diagonal parts
$j_x\left(\nu 1/2^+[660] \nu 3/2^+[651]\right)$ and
$j_x\left(\nu 3/2^+[651] \nu 5/2^+[642] \right)$ also
contribute a little to the backbendings.
From Fig.~\ref{fig4}(b) one finds that for neutron $\nu 5/2^-[523]$
the main contribution to the alignment gain after
the backbending ($\hbar \omega \sim$~0.20 MeV) comes from the
diagonal part $j_x\left(\nu 1/2^+[660]\right)$ and
the off-diagonal parts $j_x\left(\nu 1/2^+[660] \nu 3/2^+[651]\right)$ and
$j_x\left(\nu 3/2^+[651] \nu 5/2^+[642] \right)$.
Again this tells us that the
first backbendings in both cases are mainly caused by the $\nu i_{13/2}$ orbitals.
It also can be seen in Fig.~\ref{fig4}(b) that
the diagonal part $j_x\left(\nu 3/2^+[651]\right)$  increases
drastically at rotational frequency $\hbar\omega \sim 0.55$ MeV,
which means that the second backbending at the higher rotational frequency
in band 2 mainly comes from this term.
As for the alignment gain in signature $\alpha = 0$ of band 1 at higher
frequencies $\hbar\omega > 0.45$ MeV, Fig.~\ref{fig4}(c) shows that
this may comes from the diagonal parts $j_x\left(\pi 7/2^-[523]\right)$ and
$j_x\left(\pi 9/2^-[514]\right)$ and their interference term
$j_x\left(\pi 7/2^-[523] \pi 9/2^-[514]\right)$.

\begin{figure}[hbt!]
\centering
\includegraphics[width=0.8\columnwidth]{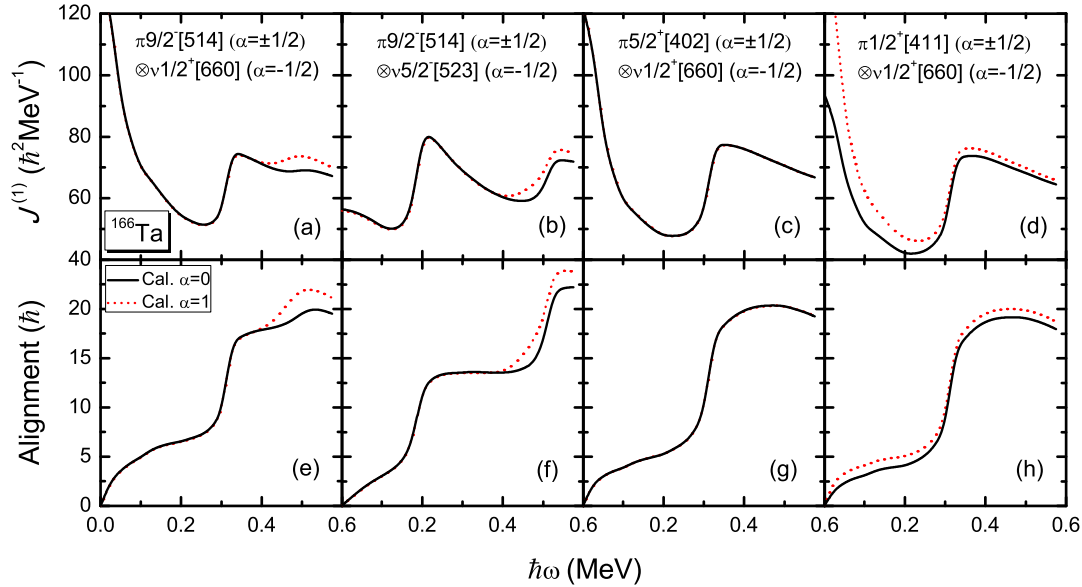}
\caption{\label{fig5} (Color online)
The same as Fig.~\ref{fig2}, but for
the calculated kinematic MOIs $J^{(1)}$ [(a)-(d)] and alignments [(e)-(h)]
for the GM partners of the observed four low-lying 2-qp bands in $^{166}$Ta.
The calculated MOIs and alignments are
denoted by black solid lines (signature $\alpha=0$,
coupled with the neutron signature $\alpha=-1/2$
and the proton signature $\alpha = + 1/2$) and red dotted lines
(signature $\alpha=1$, coupled with the neutron signature $\alpha=-1/2$
and the proton signature $\alpha =-1/2$), respectively.
}
\end{figure}

A characteristic feature of the odd-odd nucleus is the existence of the GM doublets.
When an unpaired proton and an unpaired neutron in a deformed odd-odd nucleus are coupled,
the projections of their total angular momentum on the nuclear symmetry axis,
$\Omega_{\rm p}$ and $\Omega_{\rm n}$, can produce two states with
$K_> =|\Omega_{\rm p}+ \Omega_{\rm n}|$ and $K_< =|\Omega_{\rm p} - \Omega_{\rm n}|$.
They follow the GM coupling rules~\cite{Gallagher1958_PR0111-1282}:
\begin{eqnarray}
 K_> &=& |\Omega_{\rm p} + \Omega_{\rm n}|, \
         \text{if}  \ \Omega_{\rm p}=\Lambda_{\rm p} \pm \frac{1}{2} \
         \text{and} \ \Omega_{\rm n}=\Lambda_{\rm n} \pm \frac{1}{2} \ ,   \nonumber\\
 K_< &=& |\Omega_{\rm p} - \Omega_{\rm n}|, \
         \text{if}  \ \Omega_{\rm p}=\Lambda_{\rm p} \pm \frac{1}{2} \
         \text{and} \ \Omega_{\rm n}=\Lambda_{\rm n} \mp \frac{1}{2} \ .
\nonumber
\end{eqnarray}
However, only one GM partner was observed for each configuration in $^{166}$Ta.
In the pervious calculation (Fig.~\ref{fig2}), the observed rotational bands are all
coupled from $\alpha = +1/2$ of the odd neutron with
$\alpha =\pm 1/2$ of the odd proton to form the total signature $\alpha = 1, 0$.
Therefore, it is reasonable to think that their corresponding GM partners are
coupled from $\alpha = -1/2$ of the odd neutron with
$\alpha =\pm 1/2$ of the odd proton to form the total signature $\alpha = 0, 1$.
In Fig.~\ref{fig5}, the calculated kinematic MOIs $J^{(1)}$ [(a)-(d)] and alignments [(e)-(h)]
for the GM partners of the observed four low-lying bands in $^{166}$Ta are shown.
The calculated MOIs and alignments are
denoted by black solid lines (signature $\alpha=0$,
coupled with the neutron signature $\alpha=-1/2$
and the proton signature $\alpha = + 1/2$) and red dotted lines
(signature $\alpha=1$, coupled with the neutron signature $\alpha=-1/2$
and the proton signature $\alpha =-1/2$), respectively.
Due to the large signature splitting in the neutron $\nu1/2^+[660]$ orbital,
the MOIs and alignments in band 1, band 2 and band 4 are
apparently different from their GM partners.
These calculations on kinematic MOIs and alignments
may be suggested for future experiments.

\end{multicols}

\begin{multicols}{2}

\section{Summary}{\label{sec:summ}} \vspace*{-1mm}

The experimentally observed 2-quasiparticle bands in the doubly-odd ${}^{166}$Ta are analyzed
using the cranked shell model with pairing correlations
treated by a particle-number conserving method,
in which the Pauli blocking effects are taken into account exactly.
The effective pairing interaction strength is determined by the experimental
odd-even differences in nuclear binding energies.
For each rotational band, after an appropriate Nilsson level scheme is adopted,
the experimental MOIs and alignments can be reproduced very well by the
PNC-CSM calculations, which in turn strongly support the
configuration assignments for these high-spin rotational bands.
By analyzing the occupation probability $n_\mu$ of each cranked
Nilsson orbitals near the Fermi surface and the contribution of each orbital to
the angular momentum alignments,
the mechanism of the first and second backbendings observed in these 2-qp bands
of ${}^{166}$Ta can be understood clearly.
The MOIs and alignments for the Gallagher-Moszkowski partners of these observed
high-spin rotational bands are also suggested for future experiments.

\vspace*{-2mm} \Acknowledgements{\bahao
This work is supported by the Fundamental Research Funds for the Central Universities (2015QN21)
and the National Natural Science Foundation of China (Grants No. 11275098, 11275248, 11505058).
}

%\bibliography{../../../../../Refecences/Papers}

\end{multicols}

\end{document}